\documentclass[prb,twocolumn,tightenlines,floatfix,showpacs]{revtex4}
\usepackage{graphicx}
\usepackage{amssymb}
\newcommand{\beq}{\begin{equation}}
\newcommand{\eeq}{\end{equation}}
\newcommand{\bea}{\begin{eqnarray}}
\newcommand{\eea}{\end{eqnarray}}

\begin{document}

\title{Persistence length of a polyelectrolyte in salty water:
a Monte-Carlo study}

\author{T. T. Nguyen and B. I. Shklovskii}

\affiliation{Theoretical Physics Institute, University of Minnesota, 116
  Church Street Southeast, Minneapolis, Minnesota 55455}

\begin{abstract}
We address the long standing problem of the dependence of
the electrostatic
persistence length $l_e$ of a flexible polyelectrolyte (PE)
on the screening length $r_s$ of the solution
within the linear Debye-H\"{u}ckel theory. 
The standard Odijk, Skolnick and Fixman (OSF)
theory suggests  $l_e \propto r_s^2$, while
some variational theories and computer simulations
suggest  $l_e \propto r_s$.
In this paper, we use Monte-Carlo simulations
to study the conformation of a simple polyelectrolyte. 
Using four times longer PEs than in previous simulations
and refined methods for the treatment of
the simulation data, we show that the results are consistent
with the OSF dependence $l_e \propto r_s^2$.
The linear charge density of the PE which enters
in the coefficient of this dependence
is properly renormalized to take into account local
fluctuations. 
\end{abstract}


\maketitle

\section{Introduction}
Despite numerous theoretical studies of 
polyelectrolyte (PE),
due to the long range nature
of the Coulomb interaction, the description 
of their conformation is still not as satisfactory as
that of neutral polymers. 
One of the longest standing problem is related to 
the electrostatic effect on the rigidity of
a PE. In a water solution with monovalent ions,
within the Debye-H\"{u}ckel linear screening theory,
the electrostatic interaction between PE charged monomers
has the form:
\beq
V(r)=\frac{e^2}{Dr}\exp\left(-\frac{r}{r_s}\right)~,
\label{VDH}
\eeq
where $r$ is the distance
between monomers,
$D$ is the dielectric 
constant of water, $e$ is the elementary charge,
and $r_s$ is the Debye-H\"{u}ckel screening length,
which is related to
the ionic strength $I$ of the solution by $r_s^2=4\pi l_BI$.
($l_B=e^2/Dk_BT$ is the Bjerrum length,
$T$ is the temperature of the solution).

The rigidity of a polymer is usually characterized by
one parameter, the so called persistence length
$l_p$.
For a polyelectrolyte chain, besides the intrinsic
persistence length $l_0$ which results from the specific
chemical structure of the monomers and bonds
between them, the total 
persistence length also includes an ``electrostatic"
contribution $l_e$ which results from the
screened Coulomb interactions between monomers:
\beq
l_p=l_0+l_e~.
\eeq

Because the interaction (\ref{VDH}) is exponentially
screened at distances larger than $r_s$,
early works concerning the structure of the PE 
assumed that $l_e$ is of the order of $r_s$.
However, this simple assumption
was challenged by the pioneering
works of Odijk\cite{Odijk} and
Skolnick and Fixman\cite{SF} (OSF),
who showed that Debye-H\"{u}ckel interaction
can induce a rod-like conformation at length scales much larger
than $r_s$. Their calculation gives
\beq
l_e=l_{OSF}
	=\frac{\eta_0^2}{4Dk_BT}r_s^2~,
\label{eq:lOSF}
\eeq
where 
$\eta_0$ is the linear
charge density of the PE.
Because $l_e \propto r_s^2$, it can
be much larger than $r_s$ at weak screening (large $r_s$).

Although the idea that electrostatic interaction enhances
the stiffness of a PE is qualitatively accepted and confirmed
in many experiments,
the quadratic dependence of $l_e$ on
the screening length $r_s$ is still the subject of many 
discussions. In the work of OSF, the bond angle deflection
was assumed to be small everywhere along the chain,
what is valid for large $l_0$. They suggested that
if $l_0$ is not small but $l_e$ is large enough (week screening),
their assumption is still valid.
Ref. \onlinecite{BJ}, however, has questioned this assumption 
especially when $l_0$ is so small
that the bond angle deflection is large before
electrostatics comes into play and rigidifies the
chain.

A significant progress was made by
Khokhlov and Khachaturian (KK) who proposed a 
generalized OSF theory\cite{KK}
for the case of flexible polyelectrolyte (small $l_0$).
It is known that in the absence of screening 
($r_s \rightarrow \infty$), the structure of
a polyelectrolyte can be conveniently described
by introducing the concept of electrostatic blobs.
A blob is a chain subunit within which
the electrostatic interaction is only a weak perturbation.
The blob
size $\xi$ is related to the number of Kuhn segments $g$
within one blob as $\xi=l_0g^{1/2}$. The condition
of weak Coulomb interaction suggest that
the electrostatic self energy of a blob, 
$(\eta_0gl_0)^2/D\xi$ is of the order of $k_BT$.
This leads to $\xi \simeq (Dk_BTl_0^2/\eta_0^2)^{1/3}$.
At length
scale greater than $\xi$, Coulomb interaction plays
important role and the string of blobs assumes
a rod-like conformation, with the end-to-end distance
proportional to the number of blobs.

Using this blob picture,
KK proposed that OSF theory is still applicable for
a flexible PE provided
one deals with the chain of blobs instead of the original
chain of monomers. This means, in Eq. (\ref{eq:lOSF}),
one replaces the bare linear charge density $\eta_0$
by that of the blob chain $\eta = \eta_0gl_0/\xi$.
The intrinsic persistence length $l_0$ should also be 
replaced by $\xi$.
As a result, the total persistence length of the flexible PE reads:
\beq
l_{p,KK}=\xi+\frac{\eta^2}{4Dk_BT}r_s^2~.
\label{eq:lpKK}
\eeq
Thus, in KK theory, despite the flexibility of the PE,
its electrostatic persistence length remains
quadratic in $r_s$. Small $l_0$ only
renormalizes the linear charge density
from $\eta_0$ to $\eta$.

Note that $r_s$ is implicitly
assumed to be larger than the blob size $\xi$ in KK theory 
(weak screening). For strong screening $r_s < \xi$,
there are no electrostatic rigidity and the chain behaves
as flexible chain with the Debye-H\"{u}ckel
short range interaction playing the role of 
an additional excluded volume interaction.

A number of variational
calculations have also been proposed to describe
more quantitatively the
structure of flexible chain. These
calculations, although based on different ansatz, have the
same basic idea of describing the flexible charged chain
by some model of noninteracting semiflexible chain and
variationally optimizing the persistence length of
the noninteracting system. 
Surprisingly, while some of these calculations
support the OSF-KK dependence $l_e \propto r_s^2$
such as Refs. \onlinecite{Witten,Netz,Thirumalai2},
other calculations found that $l_e$
scales linearly with $r_s$ instead\cite{BJ,Dawson,Thirumalai}.
However, because variational
calculation results depend strongly on the 
variational model Hamiltonian,
none of these results can be considered conclusive.

Computer simulations\cite{BB,Reeds,Seidel,Ullner2,Ullner3,Kremer} also 
have been used to determine
the dependence $l_e$ on $r_s$ and to verify
OSF or variational theories. Some of these papers
claim to support the linear dependence of $l_p$ on
$r_s$.  The simulation of
Ref. \onlinecite{Kremer} concludes that the
dependence of $l_p$ on $r_s$ is sublinear.
Thus, the problem of the dependence $l_e(r_s)$, despite
being very clearly stated, still remains unsolved for a
flexible PE. More details about the present status
of this problem can be found
in Ref. \onlinecite{Ullner4}.

In this paper, we again use computer simulations 
to study the dependence of $l_e$ on $r_s$.
The longest polyelectrolyte
simulated in our paper contains 4096 charged monomers,
four times
more than those studied in previous simulations.
This allows for better studying of size effect on
the simulation result. Furthermore, we use 
a more refined analysis of the simulation result, 
which takes into account local fluctuations
in the chain at short distance scale.
Our results show that OSF formula quantitatively
describes the structure of a polyelectrolyte.

The paper is organized as follows.
The procedure of Monte-Carlo simulation of
a polyelectrolyte using the primitive freely jointed
beads is described in
the next section. The data for the end-to-end distance
$R_{ee}$ is given. In Sec. III, we analyze this
data using the scaling argument to show that 
it is consistent with OSF theory. In Sec. \ref{sec:worm},
we analyze the data for the case of large $r_s$,
where excluded volume effect is not important,
in order to extract $l_e$ and again show that
it obeys OSF theory in this limit. 
In Sec. V, we use the bond angle correlation function
to calculate $l_e$ and to confirm the result
of Sec. IV. The good agreement between $l_e$ calculated
using different methods further suggests that OSF theory is
correct in describing a polyelectrolyte structure.
We conclude in Sec. \ref{sec:concl}.

Several days after the submission of our paper 
to the Los Alamos preprint
archive\cite{Nguyen}, another paper\cite{Everaers} with
Monte-Carlo simulations for PE molecules in the
same range of lengths appears in the same archive.
Results of this paper are
in good agreement with our Sec. III.

\section{Monte-Carlo simulation}
The polyelectrolyte is modeled as a chain of $N$ freely jointed
hard spherical beads
each with charge $e$.
The bond length of the PE is fixed and equal to $l_B$, 
where $l_B=e^2/Dk_BT$ is
the Bjerrum length which is about 7\AA\ at room temperature in
water solution. Thus the bare linear charge density of
our polyelectrolyte is $\eta_0=e/l_B$.
Because we are concerned about the electrostatic 
persistence length only, the bead radius is set to zero so that 
all excluded volume of monomers is provided by the screened
Coulomb interaction between them only. For convenience, the middle
bead is fixed in space. 

To relax the PE configuration globally,
the pivot algorithm\cite{pivot} is used.
In this algorithm, in an attempted move,
a part of the chain from
a randomly chosen monomer to one end of the chain is rotated
by a random angle about a random axis. This algorithm is known
to be very efficient. A new
independent sample can be produced in a computer time of the order
of $N$, or in other words, uncorrelated samples are obtained every
few Monte Carlo (MC) steps (one MC step is defined as the number
of elementary moves such that, on average, every particle
attempts to move once). 
To relax the PE configuration locally, the flip
algorithm is used. In this algorithm, a randomly chosen monomer is
rotated by a random angle about the axis connecting its two neighbor
(if it is one of the end monomers, its new position is chosen
randomly on the surface of a sphere with radius $l_B$ centered at
its neighbor.) In a simulation, the number of pivot moves is about 30\%
of the total number of moves.
The usual Metropolis algorithm is used to accept 
or reject the move. About $1\div 2\times 10^4$ MC steps are 
run for each set of parameters ($N$, $r_s$), of which
512 initial MC steps are discarded and the rest
is used for statistical average (due to time constrain, for $N=4096$, 
only 2000 MC steps are used).
Two different initial configurations, 
a Gaussian coil and a straight rod, were used to ensure 
that final states are indistinguishable and the 
systems reaches equilibrium.

\begin{figure}
\resizebox{7.5cm}{!}{\includegraphics{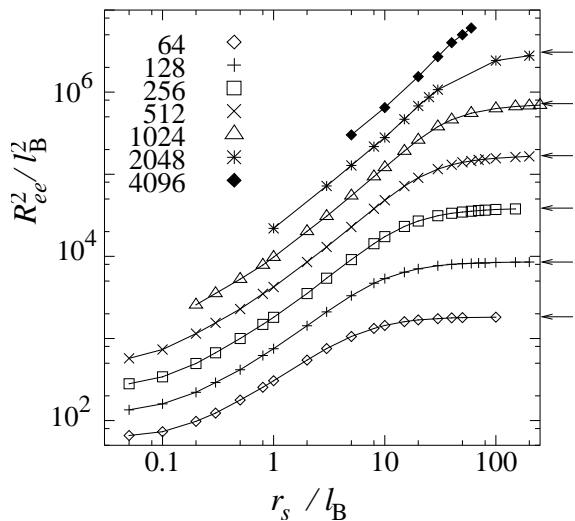}}
\caption{The square of the end-to-end distance of a 
polyelectrolyte $R_{ee}^2$ as a function of the screening 
length $r_s$ for chains with
different number of monomers $N$: 
64($\diamond$), 128($+$), 256 ($\square$), 512 ($\times$),
1024 ($\triangle$), 2048($\ast$), and 4096 ($\blacklozenge$).
The arrows on the right side show
$R_{ee}^2$ obtained 
using unscreened Coulomb potential $V(r)=e/r$.}
\label{fig:ree}
\end{figure}
The simulation result for the end-to-end distance $R_{ee}$ 
of a polyelectrolyte for different $N$
is plotted in Fig. \ref{fig:ree} as a function of the 
screening radius $r_s$ of the solution. At very small $r_s$,
Coulomb interactions between monomers are strongly screened
and the chain behaves as a neutral Gaussian
chain with $R_{ee}=l_B\sqrt{N-1}$.
At very large $r_s \gg N$, Coulomb interactions between
the monomers
are not screened and $R_{ee}$ is saturated and equal to that
of an unscreened PE with the same number of monomers
(see the arrows in Fig.  \ref{fig:ree}).

Three different methods are used to verify the validity
of OSF theory for flexible PE: 
i) study of the scaling dependence of $R_{ee}$ on $r_s$
in whole range of $r_s$, ii) extraction 
of $l_e$ in the large $r_s$ limit and
iii) analysis of the bond correlation function. In the next three
sections, we discuss these methods in details together with
their limitations. Comparison with
previous simulations is also made to explain
their results which so far have not 
supported either of the theories.

\section{Scaling dependence of $R_{ee}$ on $r_s$}

Let us first describe theoretically
how the chain size should behave as a function
of the screening radius $r_s$ when $r_s$ increases
from 0 to $\infty$.

When $r_s \ll l_B$, the Coulomb interaction is strongly screened.
Because there are no other interaction present
in our chain model of freely jointed beads, 
the chain statistic is Gaussian. Its
end-to-end distance $R_{ee}$ is proportional to the
the square root of the number of bonds and independent on $r_s$:
\beq
R_{ee}^2=l_B^2 N~.
	\label{eq:scale1}
\eeq

When $r_s \gg l_B$, the chain persistence length
is dominated by the Coulomb contribution $l_p \simeq l_e$.
If $N$ is very large such that the chain contour length
$Nl_B$ is much larger than $l_e$ then
the chain behaves as a linear chain with $Nl_B/l_e$ segments
of length $l_e$ each and thickness $r_s$.
The excluded volume between segments is
$v\simeq l_e^2r_s$, and the end-to-end distance\cite{KK}:
\beq
R_{ee}^2=l_e^2
	\left(\frac{v}{l_e^3}\right)^{2/5}
	\left(\frac{Nl_B}{l_e}\right)^{6/5}
	\propto 
	\left\{
		\begin{array}{ll}
		r_s^{4/5} & \mbox{if\ } l_e\propto r_s \\
		r_s^{6/5} & \mbox{if\ } l_e\propto r_s^2
		\end{array}
	\right.~.
	\label{eq:scale2}
\eeq

At larger $r_s$ where $l_e$ becomes comparable to the PE contour
length, the excluded volume effect
is not important. In this case, the chain statistics is again
Gaussian and
\beq
R_{ee}^2 \simeq l_e^2 
	\frac{Nl_B}{l_e}
        \propto
        \left\{
                \begin{array}{ll}
                r_s & \mbox{if\ } l_e\propto r_s \\
                r_s^2 & \mbox{if\ } l_e\propto r_s^2
                \end{array}
        \right.	~~.							
	\label{eq:scale3}
\eeq

Finally, at even larger $r_s$ when $l_p$ is greater than
$Nl_B$, the chain becomes a straight rod with length
independent on $r_s$:
\beq
R_{ee}^2\simeq l_B^2N^2~.
	\label{eq:scale4}
\eeq

If $l_e \propto r_s^2$, the transition from
the scaling range of Eq. (\ref{eq:scale2}) to 
Eq. (\ref{eq:scale3})
happens at $r_s \simeq l_BN^{1/4}$, while
the transition from
the scaling range of Eq. (\ref{eq:scale3}) to
Eq. (\ref{eq:scale4}) happens at 
$r_s \simeq l_BN^{1/2}$.
On the other hand,
if $l_e \propto r_s$, both transitions from
the scaling range of Eq. (\ref{eq:scale2}) to
Eq. (\ref{eq:scale3}) and from
the scaling range of Eq. (\ref{eq:scale3}) to
Eq. (\ref{eq:scale4}) happen at $r_s \simeq l_B N$.
This means, there is no scaling range 
of Eq. (\ref{eq:scale3}) in this theory.

\begin{figure}
\resizebox{8.5cm}{!}{\includegraphics{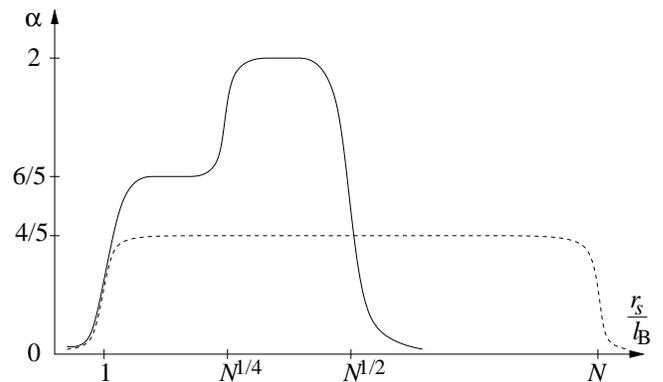}}
\caption{Schematic plot of 
$\alpha$
as a function of $r_s$ for the OSF theory $l_p\propto r_s^2$ 
(solid line) and for variational theories $l_p \propto r_s$
(dashed line).}
\label{fig:alpha}
\end{figure}

Thus, one can distinguish between the OSF result,
$l_p \propto r_s^2$, and the variational result,
$l_p \propto r_s$ by plotting the exponent
$\alpha=\partial \ln [R_{ee}^2]/\partial\ln r_s$
as a function of $\ln r_s$. The schematic figure of this
plot is shown in Fig. \ref{fig:alpha}.
OSF theory gives plateaus
at $\alpha=6/5$ and 2, and when $r_s > l_BN^{1/2}$,
$\alpha$ drops back to 0. 
Variational theories, on the other hand, would
suggest one large plateau at $\alpha=4/5$
up to $r_s \simeq l_BN$.
\begin{figure}
\resizebox{7.5cm}{!}{\includegraphics{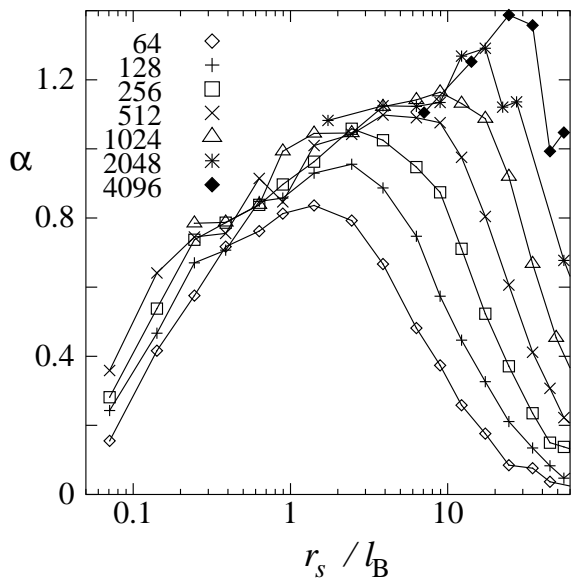}}
\caption{Simulation result for
$\alpha$
as a function of $r_s$ for different $N$:
64($\diamond$), 128($+$), 256 ($\square$), 512 ($\times$),
1024 ($\triangle$), 2048($\ast$), and 4096 ($\blacklozenge$). 
They agree reasonably well
with the solid curve of Fig. \ref{fig:alpha},
suggesting that OSF theory is correct.}
\label{fig:ree2}
\end{figure}

The simulation results for $\alpha$
are shown in Fig. \ref{fig:ree2} for different $N$.
One can see that as $N$ increases, the agreement with
OSF theory becomes more visible.
Note that
the plateaus in Fig. \ref{fig:alpha} are scaling ranges, 
and relatively sharp transitions between plateaus
are valid only for $N\rightarrow \infty$. 
For a finite $N$, the plateaus
may be too narrow to be observed and can be masked in
the transition regions. This explains why one cannot
see the plateau at $\alpha=2$ in our results. Nevertheless,
the tendencies of $\alpha$ to develop a plateau
at $\alpha=6/5$, then to grow higher toward $\alpha=2$
at larger $r_s$  and
finally to collapse to zero when approaching
relatively small $r_s=l_B\sqrt{N}$
are clearly seen for large $N$.
Thus, generally speaking, 
the curves agree with OSF theory much better than
with variational theories (where $\alpha$ is supposed
to be about 4/5 and to decrease to zero only when 
$r_s \rightarrow l_BN$, i.e. at much larger $r_s$
than what observed).

Fig. \ref{fig:ree2} also shows one reason
why similar simulations done
by other groups do not support OSF theory. All
of these simulations are limited to 512 charges.
As one can see from Fig. \ref{fig:ree2}, the curves
for $N \le 512$ do not permit to discriminate
between the two theories as clearly as
the case $N=2048$ or 4096. Only when $N$ becomes very 
large can scaling ranges with $\alpha > 1$
show up and one observes better agreement
with OSF result.

\section{Large $r_s$ limit}

\label{sec:worm}

In this section, we attempt to extract directly 
from the simulation data the
persistence length in order to compare with OSF theory.
To do this, one notices
that
even a chain with excluded volume interaction
behaves as a Gaussian chain when
its contour length is very short such that it contains only a few
Kuhn segments. In this case, one can use
the Bresler-Frenkel formula\cite{Landau} to describe the
relationship between the end-to-end distance $R_{ee}$ and
the chain persistence length $l_p$:
\beq
R_{ee}^2=2Ll_p-2l_p^2[1-\exp(-L/l_p)]~,
\label{eq:master}
\eeq
where $L$ is the contour length of the chain.

For our polyelectrolyte, this formula can be used 
for large $r_s$ when
the persistence length is of the order of $R_{ee}$
or larger. However, 
one cannot use the bare contour length $L_0=(N-1)l_B$ in
the Eq. (\ref{eq:master}) because the chain where OSF theory
is supposed to be applicable is not
the bare chain but an effective chain which takes into account
local fluctuations. The contour length $L$ of this
effective chain is
\beq
L = Ne/\eta
\label{eq:L}
\eeq
where $\eta$ is the renormalized linear charge
density of the PE. 

In KK theory, the effective chain is
the chain of electrostatic blobs,
and the normalized charge density is $\eta=\eta_0gl_0/\xi$.
However, the standard blob picture
can only be used to describe flexible 
weakly charged chains where the fraction of
charged monomers is small so that the number of monomers, $g$,
within one blob is large and Gaussian statistics can be used to
relate its size and molecular weight. 
Because, for a given number
of charged monomers, Monte-Carlo simulation for weakly 
charged polyelectrolyte is extremely time consuming,
all monomers of our simulated
polyelectrolyte are charged.
In this case, the neighbor-neighbor monomers
interaction equals $k_BT$. This makes $g\simeq 1$ and the standard 
picture of Gaussian blobs does not apply. Thus, 
in order to treat our data, we assume that
both $l_p$ and $\eta$ are unknown quantities.

To proceed further, one needs an equation
relating $\eta$ and $l_p$, and 
in order to verify OSF theory, 
we could use their formula
\beq
l_p=\eta^2r_s^2/4Dk_BT~,
\label{eq:lOSF11}
\eeq
for this purpose.
Thus, we could substitute Eq. (\ref{eq:L}) and
(\ref{eq:lOSF11}) into Eq. (\ref{eq:master}),
and solve for $\eta$ using $R_{ee}^2$ 
obtained from simulation. 
If OSF theory is valid,
the obtained values of $\eta$ should be a very slow 
changing function of $r_s$.
In addition, 
in the limit $N\rightarrow\infty$, they should
also be independent on $N$.

The OSF equation (\ref{eq:lOSF}), however, was
derived for the case $r_s\ll L$ while 
in our simulation, the
ratio $r_s/L$ is not always small. 
Therefore, instead of Eq. (\ref{eq:lOSF11}),
we use the more general Odijk's finite size
formula\cite{Odijk}
\beq
l_p=\frac{\eta^2r_s^2}{12Dk_BT}\left[
	3-\frac{8r_s}{L}+\left(5+\frac{L}{r_s}+
	\frac{8r_s}{L}
	\right)e^{-L/r_s}
	\right]~.
\label{eq:lOSF2}
\eeq
for the persistence length $l_p$.
When $L\gg r_s$, the term in the square brackets
is equal to 3 and the standard OSF result is recovered. 
On the other hand, when $r_s \gg L $, the 
persistence length $l_p$ saturates
at $\eta^2L^2/72Dk_BT$.

\begin{figure}
\resizebox{7.5cm}{!}{\includegraphics{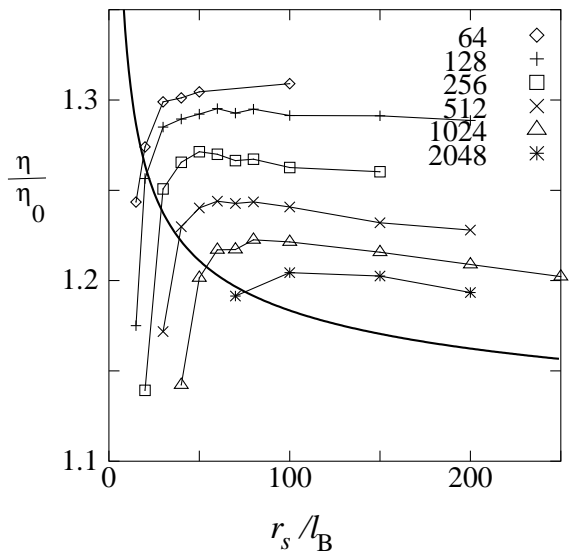}}
\caption{The linear charge density $\eta$ as
a function of the screening length for different
$N$:
64($\diamond$), 128($+$), 256 ($\square$), 512 ($\times$),
1024 ($\triangle$) and 2048($\ast$). The thick solid line
is the theoretical estimate which is the numerical
solution to Eq. (\ref{eq:F}), (\ref{eq:cost})
and (\ref{eq:etat}).}
\label{fig:eta}
\end{figure}

Below, we treat our Monte-Carlo simulation data
with the help of Eq. (\ref{eq:master})
using Eq. (\ref{eq:L})
and (\ref{eq:lOSF2}) for $L$ and
$l_p$.
The results for $\eta$ are plotted in 
Fig. \ref{fig:eta} for different PE sizes $N$.
As one can see, at large $r_s$,
$\eta$ changes very slowly with $r_s$,
and as $N$ increases, tends to saturate at
an $N$ independent value. 

It should be noted that the lines $\eta(r_s)$
in Fig. \ref{fig:eta} 
unphysically start to drop below
certain values of $r_s$. This is because at smaller
$r_s$, the electrostatics-induced excluded volume 
interactions 
between monomers become so strong that the right
hand side of Eq. (\ref{eq:master}) (which is
derived for a Gaussian worm like chain) strongly
underestimates $R_{ee}$.

Even though the picture of Gaussian blobs
does not work for our chain,
$\eta$ can still be calculated analytically
in the limit $L \gg r_s$ ($N\rightarrow \infty$).
Indeed, let us assume that 
the effective chain is straight at length scale smaller than $r_s$
(which is a reasonable assumption because all the analytical
theories so far suggested that the PE persistence length
scales as $r_s$ or $r_s^2$). Thus, the self energy of the chain
can be written as $E=L\eta^2\ln(r_s/l_B)/D$. 
At length scale smaller than $r_s$,
the polyelectrolyte behaves as a neutral chain under 
an uniform tension 
\beq
F=\partial E/\partial L=\eta^2\ln(r_s/l_B)/D~.
\label{eq:F}
\eeq
The average
angle a bond vector makes with respect to the
axis of the chain, therefore, is:
\bea
\left<\cos\theta\right>&=&\frac{\int_0^{\pi}\exp(Fl_B\cos\theta/k_BT)
	\cos\theta\sin\theta d\theta}
{\int_0^{\pi}\exp(Fl_B\cos\theta/k_BT)\sin\theta d\theta}\nonumber\\
&=&\coth\frac{Fl_B}{k_BT}-\frac{k_BT}{Fl_B}~.
\label{eq:cost}
\eea
The charge density $\eta$ can be calculated as
\beq
\eta=\eta_0/\left<\cos\theta\right>~.
\label{eq:etat}
\eeq
At weak screening $r_s \gg l_B$, it can be estimated
analytically:
\beq
\eta
\simeq\eta_0\left[1+\frac{1}{\ln(r_s/l_B)}+...\right]~,
\eeq
where the expansion terms of the order of $1/\ln^2(r_s/l_B)$ and
higher were neglected.

The more accurate numerical solution of Eq. (\ref{eq:F}),
(\ref{eq:cost}) and (\ref{eq:etat}) for $\eta$ is
plotted in Fig. \ref{fig:eta} by the thick solid line.
One can see that the values $\eta(r_s)$
calculated experimentally using OSF theory
with growing $N$ converge well to the theoretical
curve for $N=\infty$.
Remarkably, the theoretical estimate for $\eta$ 
does not use any fitting 
parameters. This, once again, strongly
suggests the OSF theory is valid for flexible 
PE as well.

The Bresler-Frenkel formula, Eq. (\ref{eq:master}),
is also used
to extract the persistence length
in Ref. \onlinecite{Kremer}
where the authors concluded that the dependence
of $l_e$ on $r_s$ is sublinear.
The authors, however, used in Eq. (\ref{eq:master})
the bare contour
length $L$, or in other words $\eta=\eta_0$,
for the calculation of $l_e$. As one
can see from Fig. \ref{fig:eta}, this leads
to 20-30\% overestimation of the contour length
of the effective chain where OSF theory is supposed
to apply. To show that this overestimation is
crucial, let us treat our data
similarly to Ref. \onlinecite{Kremer} using $\eta=\eta_0$.
We plot the resulting dependence of $l_e/r_s$ on $r_s$
(similarly to Fig. 4 of Ref.  \onlinecite{Kremer})
and compare it with our own results using corrected $\eta$.
\begin{figure}
\resizebox{7.5cm}{!}{\includegraphics{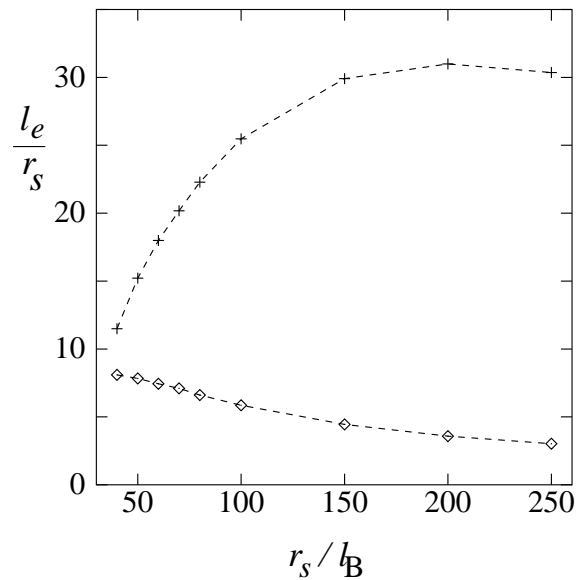}}
\caption{Plots of $l_e/r_s$ as a function of $r_s$
calculated with the help of Eq. (\ref{eq:master}),
(\ref{eq:L}) and (\ref{eq:lOSF2})
using our data for $R_{ee}^2$
($+$) and 
using unperturbed $\eta=\eta_0$ as in 
Ref. \onlinecite{Kremer}
($\diamond$). The chain with $N=1024$ is used.
}
\label{fig:lpKremer}
\end{figure}
The case $N=1024$ is shown in Fig.
\ref{fig:lpKremer}. Obviously, the two
results are different 
qualitatively. While the upper curve
follows Eq. (\ref{eq:lOSF2}) with
slightly decreasing $\eta$,
the lower curve shows sublinear growth
of $l_e$ with $r_s$ ($l_e/r_s$
is a decreasing function of $r_s$). This
sublinear dependence observed in
Ref. \onlinecite{Kremer} is clearly
a manifestation of their overestimation of 
the PE length $L$ which should be used in Eq.
(\ref{eq:master}).

Note that the true $l_e/r_s$ curve should
also eventually decrease to zero because $l_e$
saturates to the constant value $\eta^2L^2/72Dk_BT$
when $r_s \gg L$ [See Eq. (\ref{eq:lOSF2})]. 
But, according to Eq. (\ref{eq:lOSF2}), this decay
starts only at
very large $r_s$ where $r_s/L\simeq 0.25$.
The deviation from $l_e\propto r_s^2$ at large $r_s$
seen in Fig. \ref{fig:lpKremer}
is due to both the violation of the inequality
$r_s \ll L$ and
to the slight decrease of $\eta$ with $r_s$.

\section{Bond angle correlation function}

Another standard procedure used in literature is 
to calculate the persistence length
of a polyelectrolyte
as the typical decay length of the bond angle 
correlation function (BACF) along the
contour of the chain, assuming the later is exponential.
\beq
f(|s^\prime-s|)=\left<
	\cos[\angle({\mathbf b}_s,{\mathbf b}_{s^\prime})]
		\right>
\propto \exp\left(-\frac{|s^\prime-s|}{l_p}\right)~.
\label{eq:bac}
\eeq
Here ${\mathbf b}_s$ and ${\mathbf b}_{s^\prime}$
are the bond numbered $s$ and $s^\prime$ respectively
and $\angle({\mathbf b}_s,{\mathbf b}_{s^\prime})$
is the angle between them. The symbol $\left<...\right>$ 
denotes the averaging over different chain conformations.
To improve averaging, 
the pair $s$ and $s^\prime$ are also allowed to move
along the chain keeping $|s^\prime-s|$ constant. 

We argue in this section that this method
of determining persistence length actually
has a very limited range of applicability.
At either small or large $r_s$, the results
of persistence length obtained from BACF
are not reliable.
In the range where this method is supposed
to be applicable, we show that the
obtained $l_p$ are close to those obtained
in Sec. \ref{sec:worm} above.

For small $r_s$, excluded volume plays important
role and, strictly speaking, it is not
clear whether BACF is exponential, and if yes,
how one should eliminate excluded volume
effect and extract $l_p$ from the decay length.
According to Ref. \onlinecite{Ullner3}, the
decay is not exponential in this regime.

\begin{figure}
\resizebox{7cm}{!}{\includegraphics{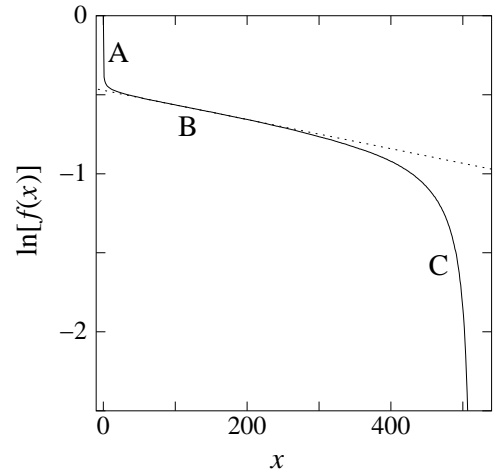}}
\caption{The logarithm of the bond correlation
function $f(x)$ as a function of the distance $x$
(in units of $l_B$)
along the chain for the case $N=512$, $r_s=50l_B$. 
There are three regions A, B and C. The dotted
line, $-0.47-x/1083$, is a linear fit of region B
suggesting that the persistence length for this
case is $l_p=1083l_B$.}
\label{fig:decay}
\end{figure}

The procedure of determining the persistence length
using BACF becomes unreliable
at large $r_s$ as well. To elaborate this point,
in Fig. \ref{fig:decay}a, 
we plot the logarithm of
the bond angle correlation function $f(x)$ along the
PE contour length for a $N=512$ and $r_s=50l_B$, 
typical values of $N$ and $r_s$ where
the excluded volume due to Coulomb interactions is small. 
There are three regions in this plot.
In region A at very small distance along the PE contour
length, monomers are within one electrostatic blobs
from each other and the effects of Coulomb interaction are small.
The bond angle correlation in this region decays
over one bond length $l_B$. At larger distance along
the PE contour length, the region B, the decay is 
exponential and a constant decay length seems well-defined.
Finally, at distance comparable to the chain's contour length, 
one again observes a fast drop of the BACF
(region C). This end effect is due to the fact that the
stress at the end of the chain goes to zero and the end bonds
become uncorrelated. 
The persistence length of interest
can be defined as the decay length in region B.

Problem arises, however, at large enough $r_s$ when
the region C (the end effect) becomes so large that
region B is not well defined. In this case the obtained
decay length underestimates the correct persistence
length. As one can
see from Fig. \ref{fig:decay}, region C can be quite
large. It occupies 40\% of the available range
of $x$, even though the screening length is only 10\% of
the contour length in this case. 

There is an even more strict condition on how large $r_s$
is when
the method of BACF loses its reliability.
If $l_e$ is larger than $L$, the decrease of
$\ln f(x)$ in region B is less than unity. 
When this happens, an exponential decay is ambiguous. 

\begin{table}
\caption{\label{tab:le}Comparison between $l_{BACF}$ calculated
using BACF method and $\eta l_e/\eta_0$ calculated in
Sec. \ref{sec:worm}. All lengths are measured in units of $l_B$.}
\begin{ruledtabular}
\begin{tabular}{cccc}
$N$	& $r_s$ & $l_{BACF}$	& $\eta l_e/\eta_0$\\ \hline
2048	& 100	& 4590	& 3682 \\ \cline{2-4}
	& 150	& 9000	& 7484 \\ \hline
1024 	& 80	& 2535	& 2180 \\ \cline{2-4}
	& 100	& 3733	& 3111 \\ \hline
512	& 50	& 1083	& 809
\end{tabular}
\end{ruledtabular}
\end{table}

Because of all these limitations,
in this section we 
use  BACF to calculate $l_e$ only
in the very limited range of $r_s$ where
excluded volume is not important and $l_e$ is not
much larger than $L$ (the decrease in region B is greater
than 0.1). The obtained $l_{BACF}$, which is measured along
the chain contour, is compared to $\eta l_e/\eta_0$ obtained
using the Bresler-Frankel formula
in the previous subsection. (The factor $\eta/\eta_0$ is
needed because $l_{BACF}$ is measured along
the real PE contour while $l_e$ is measured 
along the renormalized PE contour.) The results are
shown in the Table \ref{tab:le}. The two persistence
lengths are within 20-25\% of each other. This reasonably
good agreement between two different methods
shows that our calculations are consistent.
It further strengthens the conclusion of two previous
sections that OSF theory is correct in describing
flexible polyelectrolytes.

\section{Conclusion}

\label{sec:concl}

In this paper, we use extensive Monte Carlo simulation to
study the dependence of the electrostatic persistence
length of a polyelectrolyte on the screening
radius of the solution. Not only did we
simulate a much longer polyelectrolyte than
those studied in previous simulations in order to
show the scaling ranges, we also used a refined
analyses which take into account local
fluctuations to calculate the persistence length.
These improvements result in
a good support for OSF theory. They also
help to explain why previous simulations
failed to support OSF theory.

In order to describe our numerical data we used
a modified OSF theory in the framework of ideas of KK.
Linear charge density  $\eta$
was corrected to allow for short range fluctuations.
In our case this is a relatively small
correction to $\eta_0$ because
we deal with a strongly charged PE. When one crosses
over to sufficiently weakly charged PE linear charge density becomes
strongly renormalized and matches KK expressions.
We confirmed that corrections of  $\eta$ do not affect $r_{s}^2$
dependence of persistence length which was predicted by OSF 
for $l_0 \ll  r_s \ll L$.
In other words,  we confirm KK idea that at large 
$r_s$ all effects of flexibility
of PE are limited to a renormalization of $\eta$.
At $r_s$ comparable to contour length $L$ we 
found a good agreement of the numerical
data with OSF formula modified for
this case [Eq. (\ref{eq:lOSF2})],
which is derived in Ref. \onlinecite{Odijk}.
Again all effects of local flexibility
are isolated in the small correction to the linear 
charge density  $\eta$.

\begin{acknowledgments}
The authors are grateful to A. Yu. Grosberg
M. Rubinstein, M. Ullner and R. Netz for useful discussions
and comments. 
This work is supported by NSF No. DMR-9985785. T.T.N. is also
supported by the Doctoral Dissertation Fellowship of the 
University of Minnesota.
\end{acknowledgments}

\end{document}